\documentclass[twocolumn,showpacs,preprintnumbers,amsmath,amssymb]{revtex4}

\usepackage{graphicx}% Include figure files
\usepackage{dcolumn}% Align table columns on decimal point
\usepackage{bm}% bold math

\newcommand{\seno}{\sin \frac{\theta}{2}}
\newcommand{\coseno}{\cos \frac{\theta}{2}}

\begin{document}

\preprint{APS/123-QED}

\title{Holonomic quantum gates: A semiconductor-based implementation}% Force line breaks with \\
\author{
Paolo Solinas,$^{*}$ Paolo Zanardi,$^{\dag}$ Nino Zangh\`{\i},$^{*}$
and Fausto Rossi$^{\dag,\ddag}$
}
\affiliation{
$^*$ Istituto Nazionale di Fisica Nucleare (INFN) and
Dipartimento di Fisica, Universit\`a di Genova,
Via Dodecaneso 33, 16146 Genova, Italy \\
$^\dag$ Institute for Scientific Interchange (ISI),
Viale Settimio Severo 65, 10133 Torino, Italy \\
$^\ddag$ Istituto Nazionale per la Fisica della Materia (INFM) and
Dipartimento di Fisica, Politecnico di Torino, Corso Duca degli
Abruzzi 24, 10129 Torino, Italy }

\begin{abstract}

We propose an implementation of holonomic (geometrical) quantum gates
by means of  semiconductor nanostructures. 
Our quantum hardware consists of semiconductor macroatoms driven by  sequences of ultrafast laser pulses
({\it all optical control}).
Our logical bits are Coulomb-correlated electron-hole pairs (excitons)
in a four-level scheme selectively addressed by laser pulses with different
polarization. A universal set of single and   two-qubit gates
is generated by adiabatic change of the Rabi frequencies
of the lasers and by exploiting the dipole coupling between excitons 

\end{abstract}

\pacs{Valid PACS appear here}% PACS, the Physics and Astronomy
\maketitle

\section{Introduction}

In the recent years the interest about quantum computation (QC) and quantum
information processing (QIP) has been restless growing.
Applications of QIP e.g., quantum cryptography and quantum teleportation,
have been proposed and verified experimentally.
In QC it has  been shown that quantum algorithms
may speed up some classically intractable problems in computer
science \cite{shor}.

Unfortunately this  power inherent to  quantum features (i.e.,
entanglement, state superposition) is  difficult to be exploited  because
quantum states are typically highly unstable : the undesired coupling with
the many degrees of freedom of the environment may lead to decoherence and
to loss of the information encoded.
Another source of error can be the imperfect control of parameters
driving the evolution of the system. This can lead to
wrong output states. To implement effective  QIP techniques these two problems 
must be faced and solved.

For the problem of decoherence, some methods have been proposed theoretically:
via error correcting codes \cite{ECC} it is possible to find errors
induced by the environment and correct them.
Other approaches propose to encode information in states that are
stable against environmental  noise \cite{3}  or to eliminate
dynamically the noise effects (\cite{4},\cite{5}).
A few quantum hardwares have been proposed for implementation of quantum gates;
e.g.: nuclear magnetic resonance \cite{chuang}, ion traps (\cite{cirac},
\cite{sorensen}, \cite{molmer}, \cite{cirac2})
semiconductor quantum dots (or macroatoms) \cite{divincenzo}, \cite{biolatti};
in each of these implementations we have different gates and different ways
of processing information.

A conceptually novel  approach is  {\it topological computation}  \cite{6},\cite{7} in which the 
gate parameters depends only on global features of the control process, being therefore insensitive to local
fluctuations. Though interesting the topological gates proposed so far are
quite difficult to realize in practice because they are based on
of non-local quantum states of many body systems with complicate interactions.

Another approach that keeps some of the global (geometrical) features of the
quantum gates and seems closer to today experimental technology,
is the the so-called {\it Holonomic Quantum Computation} (HQC) (\cite{8}, \cite{9}).
In this paper we shall analyze in a detalied manner a recent proposal for
HQC with semiconductor quantum dots \cite{paper1}.
 
We shall start by recalling the basic facts about HQC (Sect. II)
and on excitonic transitions in semiconductor macroatoms (Sect. III)
In Sect IV we will show how to encode quantum information
in excitonic state an how to realize single-qubit gates by means of laser pulses.
Two-qubit gates resorting to bi-excitonic shift are illustated in Sect V.
Sect. VI contains the conclusions and an appendix is added to improve
the  self-consistency of the paper.

\section{Quantum Holonomies}

When a quantum state undergoes an adiabatic cyclic evolution, a
nontrivial phase factor appears.
This is called {\it geometrical phase} because it only depends
on global properties, i.e, not on the path in the parameter space but
only on the swept solid angle.
If the evolving state is non-degenerate we have only an Abelian phase
(Berry phase \cite{berry}), but if it is degenerate we have a non-Abelian
operator. Then we can use it to process the quantum information encoded in the
state.

More precisely, if we have a family $\mathcal{F}$  of isodegenerate 
Hamiltonians $H(\lambda)$ depending
on $m$ dynamically controllable parameters $\lambda$, we encode the
information in a $n-$fold degenerate eigenspace $\mathcal{E}$ of an Hamiltonian 
$H(\lambda_0)$. 
Changing the $\lambda$'s and driving $H(\lambda)$ along a loop we produce 
a non-trivial transformation of the initial state 
$|\psi_0 \rangle \rightarrow U |\psi_0 \rangle$. 

These transformations, 
called {\it holonomies}, are the generalization of Berry's phase and
can be computed in terms of the Wilczek-Zee gauge connection
\cite{W-Z}:  $U(C)={\bf P} exp(\oint_C A)$ where $C$ is the loop
in the parameter space and $A=\sum_{\mu=1}^m = A_{\mu} d\lambda_{\mu}$ 
is the $u(n)-$valued connection.
If $|D_i(\lambda)\rangle $ ($i=1,...,n$) are the instantaneous 
eigenstates of $H(\lambda)$, the connection is
$(A_\mu)_{\alpha\beta}=\langle D_\alpha|{\partial}/{\partial\Omega^\mu}|
D_\beta\rangle$ ($\alpha$, $\beta=1,...,n$).

It is useful to introduce the {\it curvature} $2-$form 
$F=\sum_{\mu \nu} F_{\mu \nu} dx^{\mu} \wedge dx^{\nu}$ where
$F_{\mu\nu}=\partial_\mu A_\nu - \partial_\nu A_\mu + [A_{\mu}, A_{\nu}]$;
$F$ allow us to evaluate the dimension of the holonomy group and when this
coincides with the dimension of $U(n)$ we are able to perform universal quantum
computation with holonomies.

For computation purposes we note that
if the connection components commute $[A_\mu, A_\nu]=0$, 
the curvature reduces to $F_{\mu\nu}=\partial_\mu A_\nu - \partial_\nu A_\mu$
and we can use Stoke's theorem to compute the holonomies.
The holonomic transformation can be
calculated easier $U=exp(i \int_S F_{\mu\nu} d\lambda_\mu \wedge
d\lambda_\nu)$ and depends on the 'flux' of $F_{\mu\nu}$ through
the surface $S$ delimited by $C$.
It is now clear that holonomies are associated to geometrical features
of the parameter space.

Even if with an holonomy we can build every kind of transformation
(logical gate) it is useful to think in terms of few simple gates
that constitute  an universal set (i.e., which can be composed to
obtain any unitary operator).

Many efforts have been made to implement geometrical quantum gates
(i.e. nuclear magnetic resonance \cite{jones} or super-conducting nanocircuits
\cite{falci}) because they are believed to be fault tolerant for errors
due to an imperfect control of parameters
\cite{preskill}, \cite{ellinas}.
Non-adiabatic realizations of Berry's phase logic gates have been studied
as well \cite{15}, \cite{16},\cite{non_ad}.
More recently, schemes for the experimental implementation of  non-Abelian
holonomic gates have been proposed for atomic physics, \cite{17}
ion traps \cite{DCZ}, Josephson junctions \cite{19}, Bose-Einstein 
condensates \cite{20} and neutral atoms in cavity \cite{21}.

The basic idea 
is to have a four level $\Lambda$ system with an
excited state ($|e\rangle $)
connected to a triple degenerate space with the logical qubits
($|0\rangle $ and $|1\rangle $) and an {\it ancilla} qubit
($|a\rangle $); the three degenerate state are separately
addressed and controlled.
The effective interaction Hamiltonian describing the system is
(in interaction picture)

\begin{equation}
  H_{int}= \hbar |e\rangle (\Omega_0 \langle 0| + \Omega_1 \langle 1| + \Omega_a \langle a|) + h.c.
  \label{eq:DCZ}
\end{equation}

$H$ possesses a two degenerate states (called {\it dark states}) with $E(t)=0$ 
and two {\it bright states} with $E(t)=\pm \Omega$
($\Omega = \sqrt{|\Omega_0|^2+|\Omega_1|^2+
|\Omega_a|^2}$).
At $t=0$ we codify the logical information in one of these {\it dark states}
(i.e., $|0\rangle $ or $|1\rangle $) and then, changing the Rabi
frequencies ($\Omega_i$, $i=0, 1, a$) we perform a loop in the 
parameter space ($H(0)=H(T)$).
If the adiabatic condition is full-filled at a generic time $t$ the
state of the system will be a {\it dark state} of $H(t)$ and to
the hamiltonian loop will correspond a loop for the state vector.
Since for the adiabatic condition the excited state is never populated,
the instantaneous {\it dark state} will be a superposition of the degenerate 
states.
With this loop we produce a rotation in the degenerate space
($|0\rangle $, $|1\rangle $, $|a\rangle $) starting from a logical
qubit and passing through the {\it ancilla} qubit. At the beginning and at
the end of the cycle we have only logical bits, but after a loop a
geometrical operator is applied to them.
Since we can diagonalize (\ref{eq:DCZ}) it is easy to calculate the
connection and the holonomy associated to the loop.

We can construct two single qubit gates :
$U_1 = e^{i \phi_1 |1\rangle \langle 1|}$ ({\it selective phase shift})
 and $U_2 = e^{i \phi_2 \sigma_y}$ ($\sigma_y = i(|1\rangle \langle 0|-|0\rangle \langle 1|)$).
These two gates ($U_1$ and $U_2$) are non-commutable, so we can construct
non-Abelian holonomies since $U_1 U_2 \neq U_2 U_1$.

To obtain an universal set of gates we must introduce a two bit gate;
since these gates exploit the interaction between two qubits 
they will depend on the physical systems considered.
A common choice (\cite{DCZ}, \cite{paper1}) 
is to realize a {\it selective phase shift} gate 
$U_3 = e^{i \phi_3 |11\rangle \langle 11|}$. 

\section{Excitonic transitions}
\label{sec:transitions}

In what follows we show that if we can act on a quantum dot with coherent optical (laser)
pulses, we can produce Coulomb-correlated electron-hole pairs (excitons) and
we deal with an interaction Hamiltonian similar to the one described in (\ref{eq:DCZ}).
By changing the laser parameters along the adiabatic loop, we can
produce the same single qubit gates as in \cite{DCZ}.

In the GaAs-based III-V compounds the six electrons in the valence band are
divided in a quadruplet ($\Gamma_8$ symmetry) which corresponds to
$J_{tot}=3/2$, and a doublet ($\Gamma_7$ symmetry) which corresponds to
$J_{tot}=1/2$.
If we consider a GaAs/AlGaAs quantum dot, the confining potential
(along the $z$ growth axis) breaks the symmetry and lifts the degeneracy
\cite{collins}.
The states of the quadruplet are separated in $J_z=\pm3/2$ ({\it heavy holes})
 and $J_z=\pm 1/2$ ({\it light holes}). The $\Gamma_7$ electrons have
$J_z=\pm 1/2$.
We can rewrite the eigenstates of $J_{tot}$ and $J_z$ using  the
$|S\rangle $, $|X\rangle $ , $|Y\rangle $, $|Z\rangle $ states (the four $\Gamma$ point Bloch function,
table \ref{tab:J-Jz_functions}).

If we shine the quantum dot with a laser beam we excite an electron from
the valence band to the conduction band.
In the dipole approximation we have to calculate the amplitude transition
$\langle f|{\bf \epsilon} \cdot {\bf r} |i\rangle $ (where $\epsilon$ is the
polarization vector of the electromagnetic wave, $|i\rangle $ and $|f\rangle $ are
the initial and final state respectively).

\begin{table}
  \caption{\label{tab:J-Jz_functions} $\Gamma_6$ (conduction band),
    $\Gamma_7$, $\Gamma_8$ periodic part of Bloch function.}
\begin{ruledtabular}
\begin{tabular}{l|c|c}
  $|J_{tot}, J_z\rangle $                &  $\Psi$                                                &  $\Gamma$               \\ \hline
  $|\frac{1}{2}, \frac{1}{2}\rangle $    &  $i|S\uparrow\rangle $                                        &  $\Gamma_6$             \\ \hline
  $|\frac{1}{2}, -\frac{1}{2}\rangle $   &  $i|S\downarrow\rangle $                                      &  $\Gamma_6$             \\ \hline
  $|\frac{3}{2}, \frac{3}{2}\rangle $    &  $\frac{1}{\sqrt{2}} |(X+iY)\uparrow\rangle $                  &  $\Gamma_8$ (HH)  \\ \hline
  $|\frac{3}{2}, -\frac{3}{2}\rangle $   &  $\frac{1}{\sqrt{2}} |(X-iY)\downarrow\rangle $                &  $\Gamma_8$ (HH)  \\ \hline
  $|\frac{3}{2}, \frac{1}{2}\rangle $    &  $-\sqrt{\frac{2}{3}} |Z\uparrow\rangle + \frac{1}{\sqrt{6}} |(X+iY)\downarrow\rangle $  &  $\Gamma_8$ (LH) \\ \hline
  $|\frac{3}{2}, -\frac{1}{2}\rangle $   &  $-\sqrt{\frac{2}{3}} |Z\downarrow\rangle - \frac{1}{\sqrt{6}} |(X-iY)\uparrow\rangle $  &  $\Gamma_8$ (LH) \\ \hline
  $|\frac{1}{2}, \frac{1}{2}\rangle $    &  $\frac{1}{\sqrt{3}} |Z\uparrow\rangle + \frac{1}{\sqrt{3}} |(X+iY)\downarrow\rangle $  &  $\Gamma_7$            \\ \hline
  $|\frac{1}{2}, -\frac{1}{2}\rangle $   &  $\frac{1}{\sqrt{3}} |Z\downarrow\rangle + \frac{1}{\sqrt{3}} |(X-iY)\uparrow\rangle $  &  $\Gamma_7$            \\
  \end{tabular}
  \end{ruledtabular}
\end{table}

The only non-vanishing transition amplitudes for our calculations are
 $\langle S|x|X\rangle , \langle S|y|Y\rangle , \langle S|z|Z\rangle $.

Using this relation and table \ref{tab:J-Jz_functions} we can calculate which
transitions are allowed and which ones are forbidden.

First we note that, for states like $|(X+iY)\rangle $, we can have a transition only
using 'negative' circular polarization light
${\bf \epsilon}={\bf \epsilon_x}-i{\bf \epsilon_y}$.

\begin{eqnarray}
  \langle S|{\bf \epsilon} \cdot r |(X+iY)\rangle =
  \langle S| (x-iy) |(X+iY)\rangle =  \nonumber \\
   = \langle S|x|X\rangle  + \langle S|y|Y\rangle  =2 \langle S|x|X\rangle
\end{eqnarray}

($\langle S|x|X\rangle =\langle S|y|Y\rangle $ for the symmetry of our system).

Using ``positive'' circularly polarized light we have no transition

\begin{eqnarray}
  \langle S|{\bf \epsilon} \cdot r |(X+iY)\rangle =
  \langle S| (x+iy) |(X+iY)\rangle = \nonumber \\
  = \langle S|x|X\rangle  - \langle S|y|Y\rangle  = 0
\end{eqnarray}

The latter are called {\it polarizations selection rules} ({\it PSR}).

We have also to consider the spin wave function in the initial and
final state. If the initial state has spin up (down)
the final state must have spin up (down) ({\it spin selection
rules} ({\it SSR})).
For example:

\begin{eqnarray}
  \langle S| (x-iy) |(X+iY)\rangle  \langle \uparrow|\uparrow\rangle  \nonumber =  2 \langle S|x|X\rangle
\end{eqnarray}

\begin{eqnarray}
  \langle S| (x-iy) |(X+iY)\rangle  \langle \uparrow|\downarrow\rangle  = 0
\end{eqnarray}

\subsection{Heavy-hole transitions}

From table \ref{tab:J-Jz_functions} we have the {\it heavy hole} and the $\Gamma_6$ (conduction band) states; using 
{\it SSR} we can say that the only allowed transitions are

\begin{tabular}{lccll}
  $|\frac{3}{2}, \frac{3}{2}\rangle $ &=$\frac{1}{\sqrt{2}} |(X+iY)\uparrow\rangle $ &
  $\rightarrow$ & $|\frac{1}{2}, \frac{1}{2}\rangle $ &=$i|S\uparrow\rangle $ \\
  $ | \frac{3}{2}, -\frac{3}{2}\rangle $ &=$\frac{1}{\sqrt{2}} |(X-iY)\downarrow\rangle $ &
  $\rightarrow$ & $|\frac{1}{2}, -\frac{1}{2}\rangle $ &=$i|S\downarrow\rangle $
\end{tabular}

The first transition is produced by ``negative'' circularly polarized light
(we write the corresponding operator as $\sigma^{-}$) and the second transition
is produced by ``positive'' circularly polarized light ($\sigma^{+}$) for the
{\it PSR}.

In terms of excitons (electron-hole pairs) if we perform a transition
with $\sigma^{-}$, we promote an electron with spin $3/2$ of the valence
band to the conduction band with spin $1/2$ and we get an exciton
with angular momentum $-1$ ($E^{-}$). With $\sigma^{+}$ we promote an
electron with spin $-3/2$ of the valence band to the conduction band with
spin $1/2$ and we have an exciton with angular momentum $1$ ($E^{+}$).

\subsection{Light hole transitions}

For the {\it light hole} we have more allowed transitions; this is due to
the presence of the $|Z\rangle $ states in the wave function. As for the HH
transitions, using $\sigma^{\pm}$ we have

\begin{tabular}{lcl}
  $|\frac{3}{2}, \frac{1}{2}\rangle $ &  $\underbrace{\longrightarrow}_{\sigma^{-}}$ &
  $|\frac{1}{2}, -\frac{1}{2}\rangle $ \\
  $|\frac{3}{2}, -\frac{1}{2}\rangle $ & $\underbrace{\longrightarrow}_{\sigma^{+}}$
  & $|\frac{1}{2}, \frac{1}{2}\rangle $
\end{tabular}

These transitions are allowed with circular (positive or negative)
polarization (${\bf \epsilon}={\bf \epsilon_x} \pm i{\bf \epsilon_y}$) and
propagation along the $z$ (growth) axis.
If we have the wave propagating along the $x$ or $y$ axis and the polarization
along $z$ the transition is allowed by {\it PSR}.
Using also the {\it SSR} we get the two allowed transitions:

\begin{equation}
  \langle \frac{1}{2}, \frac{1}{2}| z  |\frac{3}{2}, \frac{1}{2}\rangle
   \sim  \langle S|z|Z\rangle
\end{equation}

\begin{equation}
  \langle \frac{1}{2}, -\frac{1}{2}| z  |\frac{3}{2}, -\frac{1}{2}\rangle
  \sim  \langle S|z|Z\rangle
\end{equation}

With the operator $\sigma^0$ we have the following transitions

\begin{tabular}{lcl}
  $|\frac{3}{2}, \frac{1}{2}\rangle $ & $\underbrace{\longrightarrow}_{\sigma^{0}} $
  & $|\frac{1}{2}, \frac{1}{2}\rangle $ \\
  $|\frac{3}{2}, -\frac{1}{2}\rangle $ & $\underbrace{\longrightarrow}_{\sigma^{0}}$
  & $|\frac{1}{2}, -\frac{1}{2}\rangle $
\end{tabular}

Such transitions with polarization along $z$ have been
experimentally observed \cite{marzin}.

Exciting  light-hole electrons with three different kinds of
light (left and right circular polarization and polarization along $z$ axis)
we can induce three different kinds of transitions with the same energy
\cite{marzin}.

In terms of excitons if we make a transition
with $\sigma^{\pm}$, we promote an electron with spin $\mp1/2$ from the
valence band to the conduction band with spin $\pm1/2$ and we get an exciton
with angular momentum $\pm1$ ($E^{\pm}$).
Using light propagating along $x$ or $y$ with $z$ polarization we promote an
electron with spin $\pm 1/2$ from the valence band to the conduction band with
spin $\pm1/2$ and we have an exciton with angular momentum $0$ ($E^{0}$).

\begin{figure}
  \includegraphics[height=4cm]{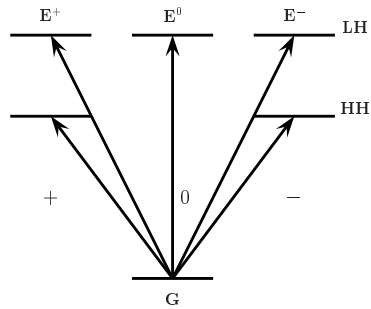}% Here is how to import EPS art
  \caption{\label{fig:lh-hh} Level scheme for LH and HH.}
\end{figure}

The allowed transitions and
the corresponding energy-level scheme for HH and LH are shown in
fig. \ref{fig:lh-hh}.

\subsection{$\Gamma_7$ transitions}

In the same way we can compute the transition selection rules for the
$\Gamma_7$ electrons.

\begin{tabular}{lcl}
  $|\frac{1}{2}, \frac{1}{2}\rangle $ &  $\underbrace{\longrightarrow}_{\sigma^{-}}$ &
  $|\frac{1}{2}, -\frac{1}{2}\rangle $ \\
  $|\frac{1}{2}, -\frac{1}{2}\rangle $ & $\underbrace{\longrightarrow}_{\sigma^{+}}$
  & $|\frac{1}{2}, \frac{1}{2}\rangle $ \\
  $|\frac{1}{2}, \frac{1}{2}\rangle $ & $\underbrace{\longrightarrow}_{\sigma^{0}} $
  & $|\frac{1}{2}, \frac{1}{2}\rangle $ \\
  $|\frac{1}{2}, -\frac{1}{2}\rangle $ & $\underbrace{\longrightarrow}_{\sigma^{0}}$
  & $|\frac{1}{2}, -\frac{1}{2}\rangle $ \\
\end{tabular}

Like for the LH, we have three different kinds of transitions that
can be distinguished by the light polarization.

Those transitions are energetically higher with respect to the
LH and HH ones. Therefore, we should be able to inibite them using
properly tuned laser sources with bandwidth
$\Delta E < E_{\Gamma_7} - E_{LH} \simeq  0.3$\,eV
\cite{bastard}.

\section{Exciton Interaction Hamiltonian and Single Qubit Gates}

Now we want to write the
interaction Hamiltonian for the exciton transitions (excluding $\Gamma_7$
transitions).

The Hamiltonian for the light-matter interaction  is (we use the electric
field instead of the vector potential  \cite{schmitt})

\begin{equation}
  H_{int}= -e [\vec{P} \cdot \vec{E}^{*}(t)+ h.c.]
  \label{eq:interac}
\end{equation}

where  $\vec{E}(t)$ is the electric field,  $\vec{P}$ is the polarization
operator defined as

\begin{equation}
  \vec{P}=\sum_{n,m} v_{m}^{\dagger} c_n \langle v,m|e \vec{r} |c,n\rangle
  =\sum_{n,m} v_{m}^{\dagger} c_n \vec{\mu}^{*}_{nm}
\end{equation}

and

\begin{equation}
   \vec{\mu}_{nm}=\langle c,n|e \vec{r} |v,m\rangle
\end{equation}

$c_n$ and $c_{n}^{\dagger}$ are the annihilation and creation operator for
an electron in the conduction band with spin $n$ ($n=\pm1/2$);
$v_m$ and $v_{m}^{\dagger}$ are the annihilation and creation operators for
an electron in the valence band with spin $m$ ($m=\pm1/2$ (LH) or
$m=\pm3/2$ (HH) ).

Then, using the dipole approximation ($\vec{E}^{*}(t)=
  E_0 e^{i({\bf k} {\bf x}- \omega t)} {\bf \epsilon} \approx
  E_0 e^{- i \omega t}{\bf \epsilon} $)
\begin{equation}
  H_{int}= -[\sum_{n,m} v_{m}^{\dagger} c_n \langle v,m|e \vec{r} |c,n\rangle  \cdot
  \vec{E}^{*}(t)+ h.c.]
  \label{eq:ham_crea_dist}
\end{equation}

We define

\begin{equation}
  \hbar \Omega_{n,m}=\vec{\mu}_{nm}^{*} \cdot \vec{E}^{*}(t)=
  E_0 e^{-i\omega t} {\bf \epsilon} \cdot \langle v,m|e \vec{r} |c,n\rangle
  \label{eq:omega}
\end{equation}

The last term is the dipole transition amplitude.

The term $c_{\pm 1/2}^{\dagger} v_{\pm 3/2}$ 
describes the promotion of an electron with spin $\pm 3/2$ to the 
conduction band with spin $\pm 1/2$ and then it describes the 
creation of an 'heavy' exciton with angular momentum $\pm1$ ($E^{\pm}$) 
from the ground state ($G$).
In the same way we can rewrite the terms in (\ref{eq:ham_crea_dist})
taking account of {\it light hole} transition.
With this new notation,
 we have non-vanishing coefficients (as discussed in
section \ref{sec:transitions}) in table \ref{tab:omega}.

\begin{table}
  \caption{\label{tab:omega} Rabi frequencies for allowed transitions.}
  \begin{ruledtabular}
    \begin{tabular}{lrcrc}
  $\Omega_{n,m}$           &  v  &   & c &  exciton \\ \hline
  $\Omega_{\frac{1}{2},\frac{3}{2}}$   & $\frac{3}{2}$ &$\longrightarrow$&   $\frac{1}{2}$ &
  $E^{-}$ \\
  $\Omega_{-\frac{1}{2},-\frac{3}{2}}$ & $-\frac{3}{2}$ & $\longrightarrow$ & $-\frac{1}{2}$ &
  $E^{+}$ \\
  $\Omega_{\frac{1}{2},-\frac{1}{2}}$  & $-\frac{1}{2}$ & $\longrightarrow$ & $ \frac{1}{2}$ &
  $E^{+}$ \\
  $\Omega_{-\frac{1}{2},\frac{1}{2}}$  & $\frac{1}{2}$  & $\longrightarrow$ & $ -\frac{1}{2}$ &
  $E^{-}$ \\
  $\Omega_{\frac{1}{2},\frac{1}{2}}$   & $\frac{1}{2}$  & $\longrightarrow$ & $  \frac{1}{2}$ &
  $E^{0}$ \\
  $\Omega_{-\frac{1}{2},-\frac{1}{2}}$ & $-\frac{1}{2}$ & $\longrightarrow$ & $ -\frac{1}{2}$ &
  $E^{0}$ \\
    \end{tabular}
  \end{ruledtabular}
\end{table}

The Hamiltonian becomes

\begin{eqnarray}
  H_{int} = - \hbar [
  \Omega_{-,\mbox{{\footnotesize HH}}}|E^{-}_H\rangle \langle G| +
  \Omega_{+,\mbox{{\footnotesize HH}}}|E^{+}_H\rangle \langle G| + \nonumber \\
  \Omega_{+,\mbox{\footnotesize LH}}|E^{+}_L\rangle \langle G| +
  \Omega_{-,\mbox{\footnotesize LH}}|E^{-}_L\rangle \langle G| + \nonumber \\
  \Omega_{0,\mbox{\footnotesize LH}}|E^{0}_L\rangle \langle G| +h.c.]
\end{eqnarray}

In the last term we include the two identical kinds of
$E^{0}$ excitons.

As we stated before, if we can address the light or heavy hole we can
distinguish between $E^{\pm}_{\mbox{{\footnotesize HH}}}$ and 
$E^{\pm}_{\mbox{{\footnotesize LH}}}$; so using light with
specified frequency tuned to LH transition, we can write:

\begin{eqnarray}
  H_{int} & = &
 - \hbar (\Omega_{+,\mbox{\footnotesize LH}}|E^{+}_L\rangle +
   \Omega_{-,\mbox{\footnotesize LH}}|E^{-}_L\rangle +
   \Omega_{0,\mbox{\footnotesize LH}}|E^{0}_L\rangle )\langle G| \nonumber \\
   & + & h.c.
\end{eqnarray}

This Hamiltonian has the same structure as the one proposed in \cite{DCZ} 
to implement the holonomic quantum computation with trapped ions.
So we can construct the same geometrical single qubit gates
($U_1$ and $U_2$) using , for example,
$E^{+}$ and $E^{-}$ as $|1\rangle $ and $|0\rangle $ bits respectively and 
$E^{0}$ as {\it ancilla} bit $|a \rangle $.

For the first gate we choose $\Omega_{-} = 0$,
$\Omega_{+} = -\Omega \sin(\theta /2) ~e^{i \varphi}$ and
$\Omega_{0} = \Omega \cos(\theta /2)$. The dark states
are given by  $|E^{-}\rangle$ and
$|\psi\rangle= \cos(\theta /2) |E^{+}\rangle + \sin(\theta /2)~ e^{i \varphi} |E^{0}\rangle
$.
By evaluating the connection  associated to this
two-dimensional degenerate eigenspace, it
is not difficult to see that    the unitary transformation
$U_1 = e^{i \phi_1 |E^{+}\rangle\langle E^{+}|}$
($\phi_1 =\frac{1}{2} \oint \sin\theta~ d\theta~d\varphi$) 
can be realized as an holonomy.
For the second gate we choose $\Omega_{-} = \Omega \sin\theta \cos\varphi$,
$\Omega_{+} = \Omega \sin\theta \sin\varphi$ and
$\Omega_{0} = \Omega \cos\theta$. The dark states are now given by
$|\psi_1\rangle = \cos\theta \cos\varphi |E^{-}\rangle + \cos\theta \sin\varphi |E^{+}\rangle - \sin\theta |E^{0}\rangle$ and
$|\psi_2\rangle = \cos\varphi |E^{+}\rangle - \sin\varphi |E^{-}\rangle$.
In this case, the unitary transformation
$U_2 = e^{i \phi_2 \sigma_y}$ (where $\phi_2 = \oint \sin\theta d\theta
d\varphi$ and $i \sigma = |E^+ \rangle \langle E^-| - |E^- \rangle \langle E^+|$)
can be implemented.

We performed numerical simulations to show how our scheme works and how we
can satisfy adiabaticity request and apply logical gates.
The exciton states have energies between $1.5~eV$ and $1.7~eV$ which 
correspond to sub-femto second time scale; then using femtosecond 
laser pulse we avoid transition between {\it ground} and exciton state
during the evolution.
Using Rabi frequencies about $0.02$~fs$^{-1}$ 
(corresponding to $\Omega^{-1} = 50$\,fs) and 
evolution times of $T_ad = 7.5$~ps (as in the simulation) we get for the 
adiabatic condition $\Omega ~T_{ad} = 150 \gg 1$ which assures us that 
there will be no transition between {\it dark} and {\it bright} states
(separated by $\Omega$ energy).

\begin{figure}
  \includegraphics[height=5cm]{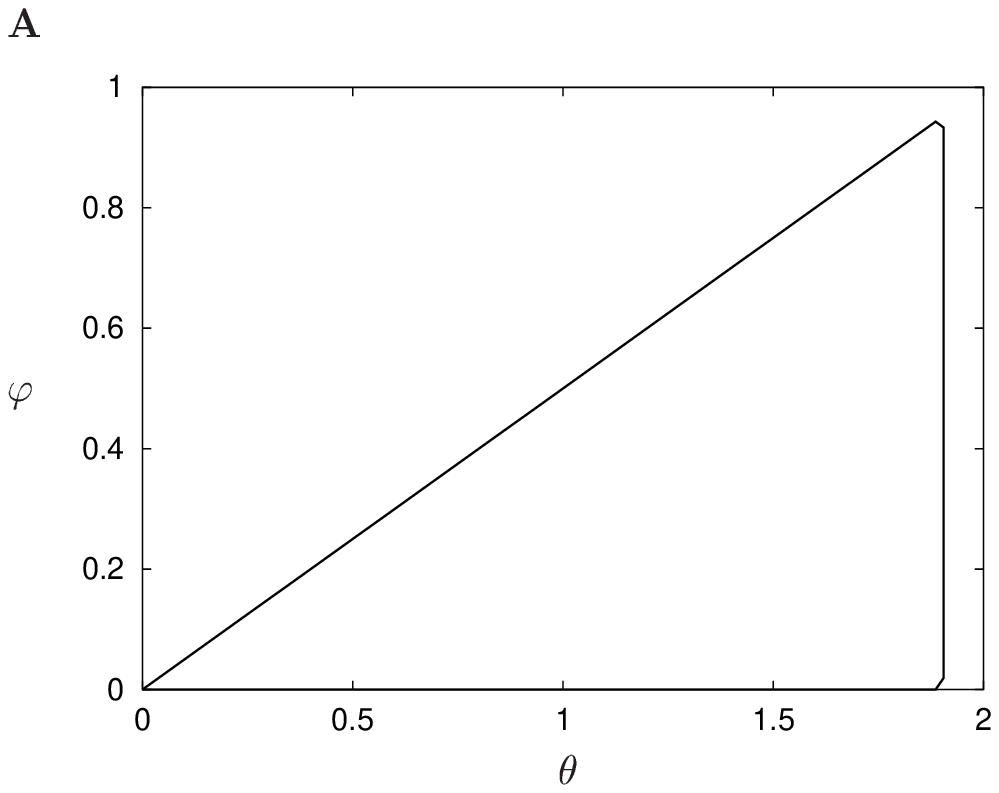} 
  \includegraphics[height=6cm]{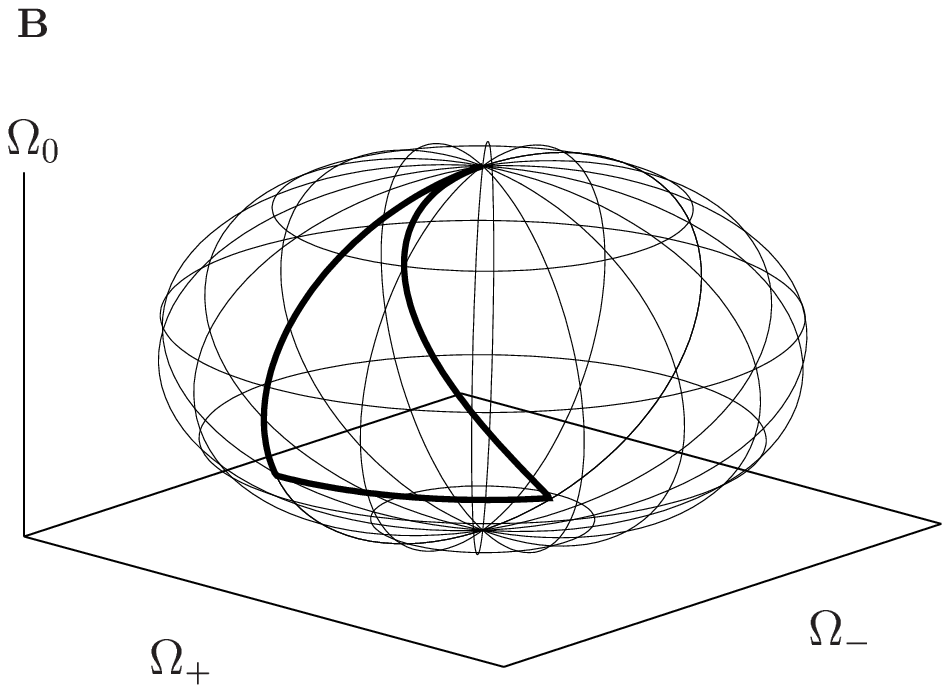} 
  \caption{\label{fig:theta-phi} (A) Loop in the $\theta - \phi$ parameter
    space ($\theta_m=\pi$). (B) Evolution of the $|E^{+}\rangle$ in the 
  $|E^{-}\rangle - |E^{+}\rangle - |E^{0}\rangle$  space for gate $2$ 
  and $\phi_2 = \pi /2$.}
\end{figure}

\begin{figure}
  \includegraphics[height=5cm]{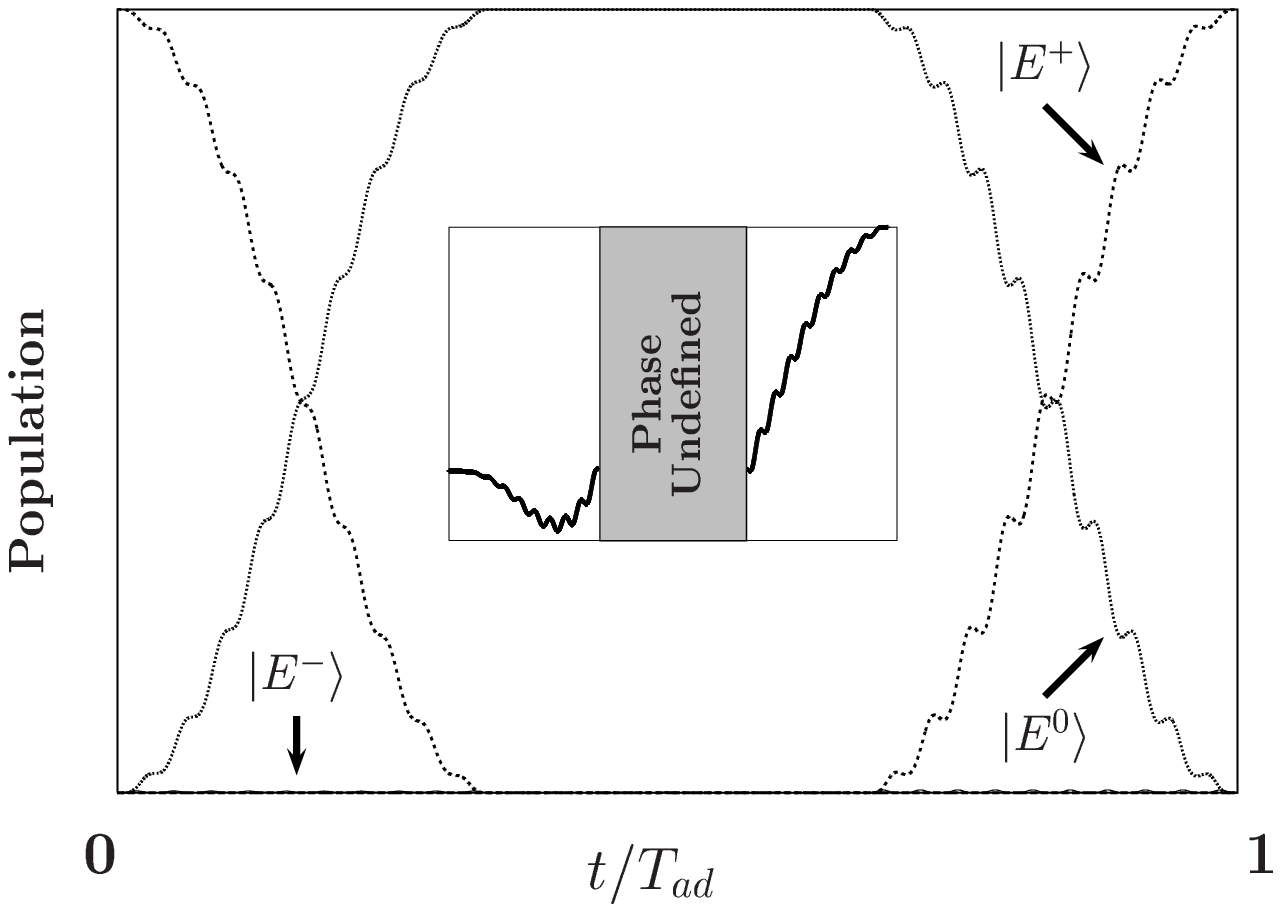}
  \includegraphics[height=5cm]{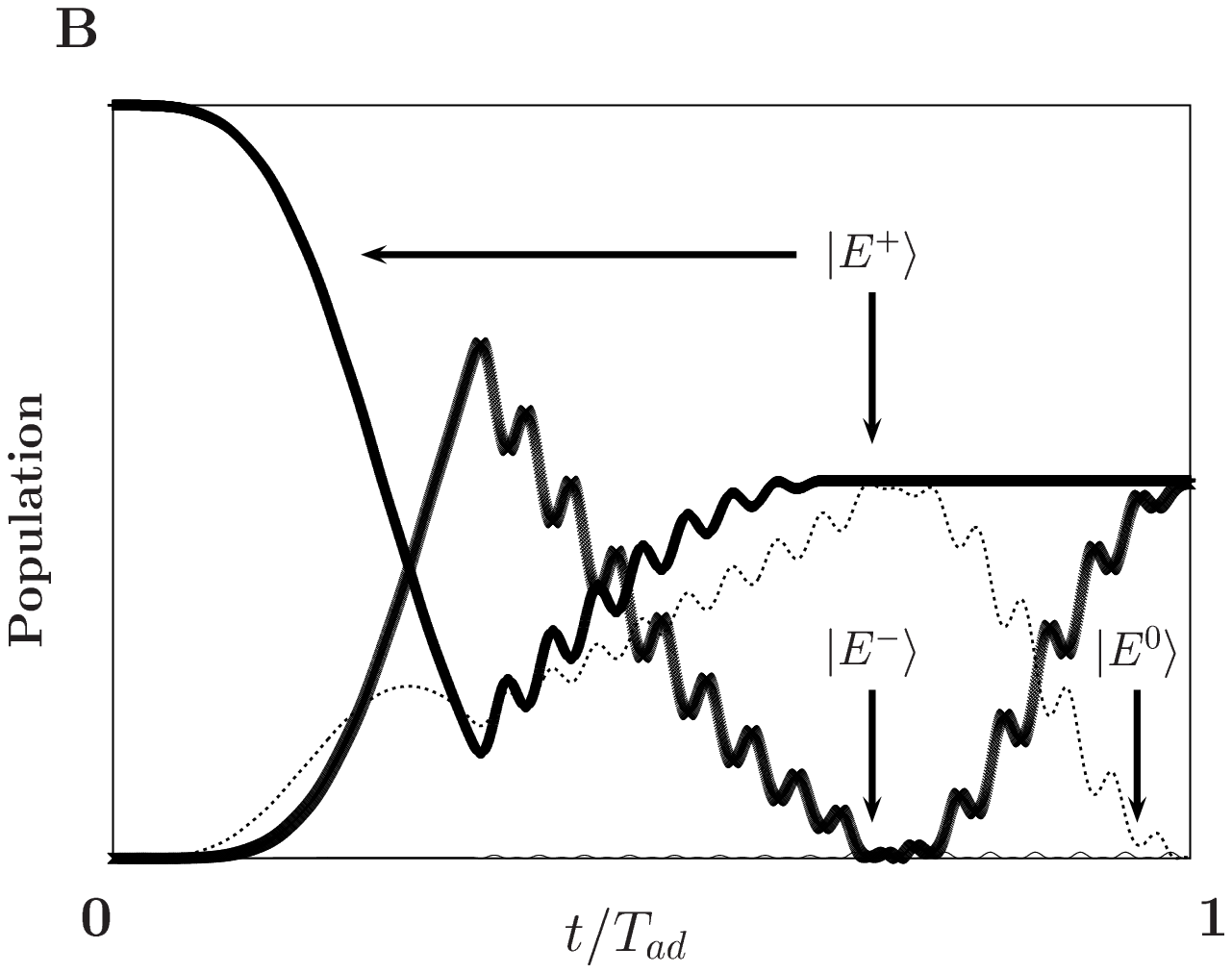}
  \caption{\label{fig:populations} (A) Simulated time evolution of the 
HQC gate
1 with $\phi_1=\pi/4$ and initial state $|E^{+}\rangle$.
The inset shows (where it is defined) the quantity $ \varphi$ where 
$ \varphi := \mbox{Arg}  \langle\Psi(t)|E^+\rangle/  |
\langle\Psi(t)|E^+\rangle|$.
(B) Simulated time evolution of the HQC gate 2 with $\phi_2=\pi/4$ 
(Hadamard gate) and initial state $|E^{+}\rangle$.}
\end{figure}

\begin{figure}
  \includegraphics[height=6cm]{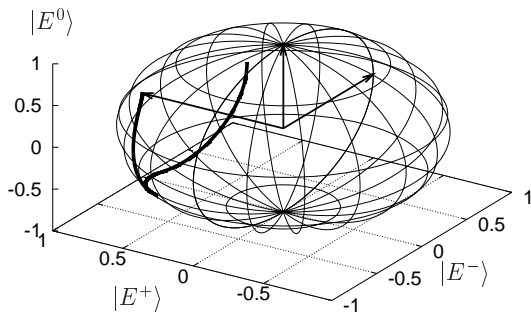} 
  \caption{\label{fig:state_loop} Evolution of the initial state 
$|E^{+}\rangle$ in the   $|E^{-}\rangle - |E^{+}\rangle - |E^{0}\rangle$  
space for gate $2$ and $\phi_2 = \pi /2$.}
\end{figure}

In Figure \ref{fig:theta-phi}(A) the loop in the $\theta-\phi$ space
is shown.
Since the holonomic operator depends on the solid angle
($\oint d\Omega = \oint d\theta d\phi \sin \theta$), the only
contribute from this loop comes from the first part 
and can be easily  calculated
$\int d\Omega = 1/2 ( \sin\theta_m - \theta_m \cos\theta_m)$.
Then it is sufficient to change $\theta_m$ to apply a different
operator.
In Fig.~\ref{fig:theta-phi} (B) we show the loop in the control parameters 
manifold for gate 2 
($\Omega^{-}$, $\Omega^{+}$, $\Omega^{0}$), since the parameters are 
real the 3D vector $\vec{\Omega}$  evolves on a sphere.
These two figures refer to the implementation of Hadamard gate 
(also shown in Fig. \ref{fig:populations} (B)  and in Fig. 
(\ref{fig:state_loop}) )
and we choose $\theta_m$ in order to obtain $\oint d\Omega = \pi/4$. 

Figure \ref{fig:populations} shows the state populations
during the quantum-mechanical evolution; as we can see, the state $|G\rangle$
is never populated (as expected in the adiabatic limit).
For the case of gate 1 [see Fig.~\ref{fig:populations}(A)]
the $|E^{-}\rangle$ state is decoupled in the evolution while
the state $|E^{+}\rangle$ evolves to the {\it ancilla} state 
($|E^{0}\rangle$), to eventually end in
$|E^{+}\rangle$ (as we expect for the dark state).
In the inset we show the phase accumulated by the $|E^+\rangle$ state;
of course, in the central region the phase is undefined.

The quantum evolution of gate 2 in Fig.~\ref{fig:populations}(B) 
is more complicated because there are not decoupled states and all
the three degenerate states are populated. 
We start from $|E^{+}\rangle$ and end in a superposition of 
 $|E^{+}\rangle- |E^{-}\rangle$.
It can be better understood by looking at Fig.~
\ref{fig:state_loop}, 
where we show the evolution of the dark state in the $|E^{+}\rangle$, 
$|E^{-}\rangle$, $|E^{0}\rangle$ space.
As mentioned above, the initial dark state evolves in the degenerate space: 
it starts from  the $|E^{+}\rangle$ axis, then  
passes through a superposition of the three states and ends in the 
$|E^{+}\rangle -|E^{-}\rangle $ plane ($(|E^{+}\rangle +|E^{-}\rangle)/
\sqrt{2}$ state).

The numerical simulations show that our scheme works and we are 
able to produce the desidered gates with realistic parameters 
for the semiconductor quantum dots \cite{29} and for the recent 
ultrafast laser technology \cite{26}.
Moreover it is clear (also with the gates in \cite{paper1}) that we are able 
to apply different gates in the same gating time because the latter  depends 
only on  the adiabatic constraint (and not on the gate we choose) and though the 
adiabatic limitation we can apply several quantum gates.
Infact recent studies \cite{30} have shown that excitons can exhibit a long 
dephasing time (comparable to hole-electron recombination time) on nanosecond 
time-scale. The degeneracy in our model has an importnt role (even if the 
request can be made weaker and we can use almost-degenerate state 
i.e. see section \ref{sec:laser}) and this can further prolong the dechoerence
time till the recombination of light-hole.

\subsection{Laser bandwidth }
\label{sec:laser}

We saw that by using light with different polarizations we can induce different
transitions and generate  $E^{\pm}$, $E^0$ excitons.
To select which electron to excite (HH, LH, $\Gamma_7$)
we have to use different energies; in fact the  $\Gamma_7$ transitions
are the most energetic, then there are the LH and  the HH.

For circular ($\pm$) polarization light propagating along the $z$ axis we have
\cite{bastard} that the ratio of probabilities to excite the
relative electron is

\begin{tabular}{ccc}
  $\frac{\mbox{\footnotesize HH}}{\mbox{\footnotesize LH}}$ & = & 3\\
  $\frac{\mbox{\footnotesize HH}}{\Gamma_7}$ & = & $\frac{3}{2}$
\end{tabular}

So it is sufficient  that the laser bandwidth is not too large
($\Delta E <  E_{\mbox{\footnotesize LH}}-E_{\mbox{\footnotesize HH}}$ but
$\Delta E \ll E_{\Gamma_7}-E_{\mbox{\footnotesize LH}}$) to  excite HH 
instead of LH and forbid the $\Gamma_7$.

For light propagating along the $x$($y$) axis with $z$ polarization the
HH transition are forbidden and

\begin{tabular}{ccc}
  $\frac{\mbox{\footnotesize LH}}{\Gamma_7}$ & = 2 &
\end{tabular}

So even if this laser bandwidth is  $\Delta E <  E_{\Gamma_7}-
E_{\mbox{\footnotesize LH}}$ it is more likely to produce LH.
As we wrote before in practical situation we should be able to prohibit 
$\Gamma_7$ transition just with this choices.

\begin{figure}
  \includegraphics[height=4cm]{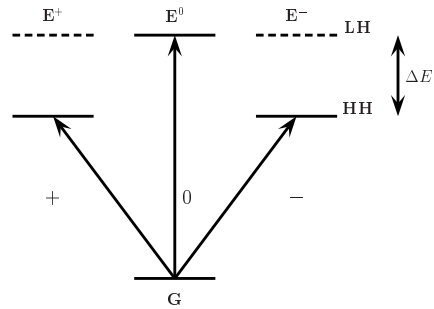}
  \caption{\label{fig:mixing} Level scheme for geometrical gates when is
    impossible to address only HH or LH.}
\end{figure}

Now we show that 
even if we are not able to energetically distinguish HH and LH 
the holonomic scheme proposed works as usually.
The level scheme for this configuration is shown in Fig \ref{fig:mixing}.
We can excite $E^{\pm}_{\mbox{\footnotesize HH}}$ excitons or 
$E^{0}_{\mbox{\footnotesize LH}}$ exciton.
If we have an adiabatic evolution fast enough, the three levels are mixed
during the evolution and so, for our scheme, they can be considered
degenerate. 

The energy gap between HH and LH excitons is of the order of
$0.05-0.03~eV$
( \cite{marzin}, \cite{miller2}, \cite{miller1}, \cite{matsumoto}), whereas
between $\Gamma_7$ and HH-LH the gaps is about $0.3~ eV$.
Both of this energy gap are very large compared to the bandwidth of the
pico and femtosecond pulsed laser, so  in practical applications
one should be able to separate LH and HH excitons.

\subsection{Dynamical phase }

During the evolution along the adiabatic loop the state acquire a
dynamical phase in addition to the geometrical phase.
In the first proposal of adiabatic gates with standard two level 
systems additional work is needed to eliminate this undesidered 
phase.
In ref.~\cite{ekert} they show how this dynamical phase can be eliminated: 
we have to run the geometrical gate several times in order to let the
dynamical phases cancel each others.
The drawback in this method is that we have to iterate several
times the adiabatic gate and, because of the adiabatic condition,
long time is needed to apply the final geometrical gate.

In this model, if we use LH excitons, the logical 
and the {\it ancilla} states are
degenerate and the ground state is never populated during the
evolution; so the dynamical phase shift is the same for
the two logical qubits and can be neglected.

If we encode logical information in the HH excitons 
($\pm$) and use the LH exciton ($0$) as {\it ancilla} qubit 
we have an energy difference $\Delta E$ and then a dynamical phase appears.
Again, we can neglect it, because at the begining (encoding of information) 
and at the end (reading information) of the evolution, the $|E^0\rangle$
state is never populated and then the phase difference does not affect 
the logical information.
Then, in both models, we can avoid problems with the dynamical phase.

\section{Two qubits gate }

For the two qubit gate we cannot take directly the DCZ model but we use the
bi-excitonic shift \cite{biolatti}.
In fact if we have two coupled quantum dots the presence  of an exciton in
one of them (e.g. in dot b) produces a shift in the energy level of the other
(e.g. dot a) from $E$ to $E + \delta/2$.

\begin{figure}[t]
  \includegraphics[height=4cm]{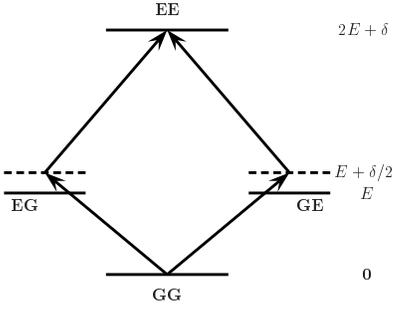}
  \caption{\label{fig:two_ph} Level scheme for the two-photon process.}
\end{figure}

Let's consider the two dots in the ground state $|G G\rangle $;
if we shine them with circular ('positive' or 'negative')
light at $E+\delta/2$ energy we should be able to produce two
excitons $|E E \rangle $ (see appendix \ref{app:two_ph}).
For energy conservation this is the only possible transition
(the absorption of a single photon is at energy $E$).
The detuning allows us to isolate the two-exciton space
($|E E \rangle $) from the single exciton space
($|E G \rangle $, $|G E \rangle $). 
The level scheme is shown in Fig. \ref{fig:two_ph}.

\begin{figure}
  \includegraphics[height=5cm]{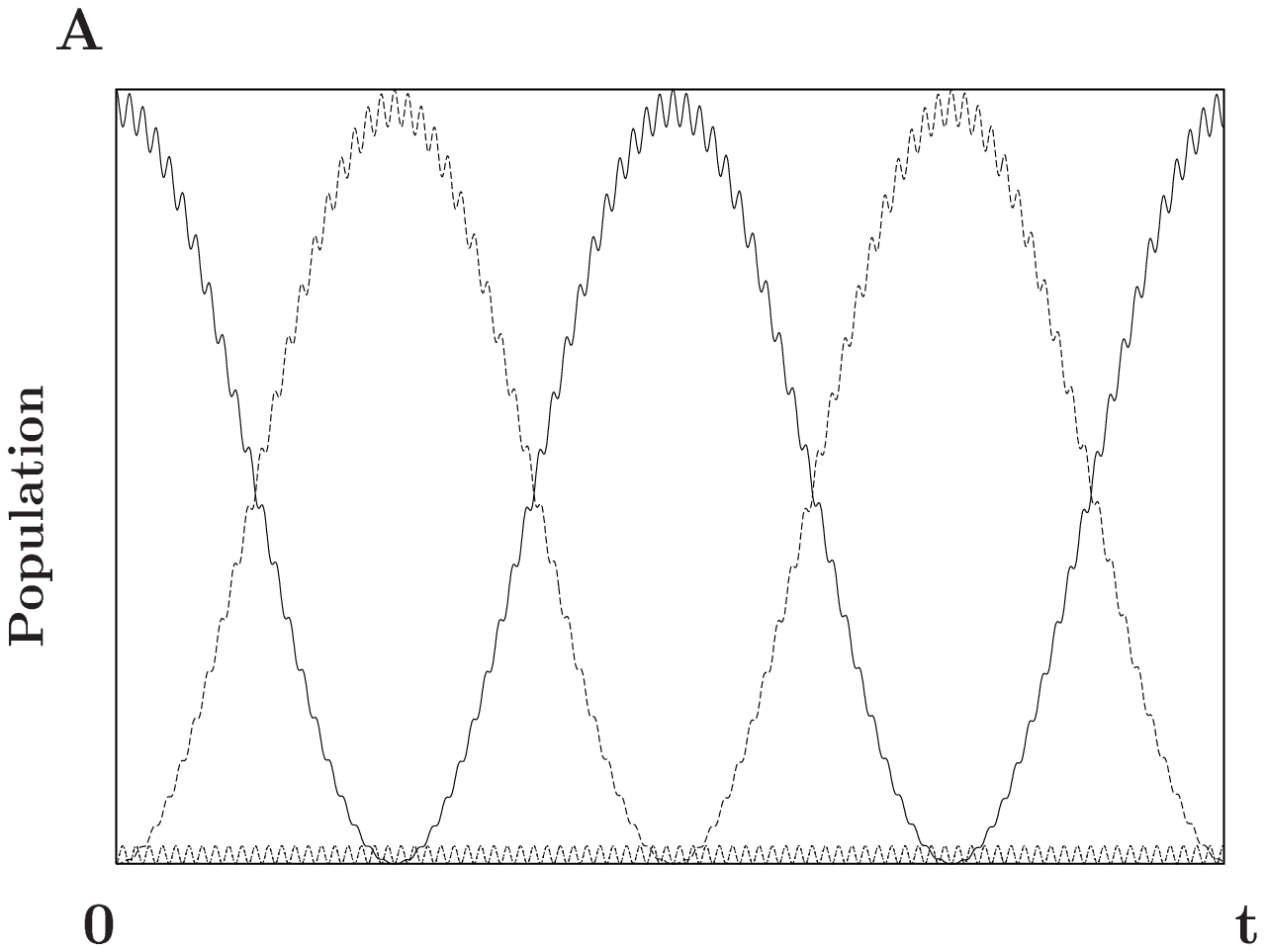}
  \includegraphics[height=5cm]{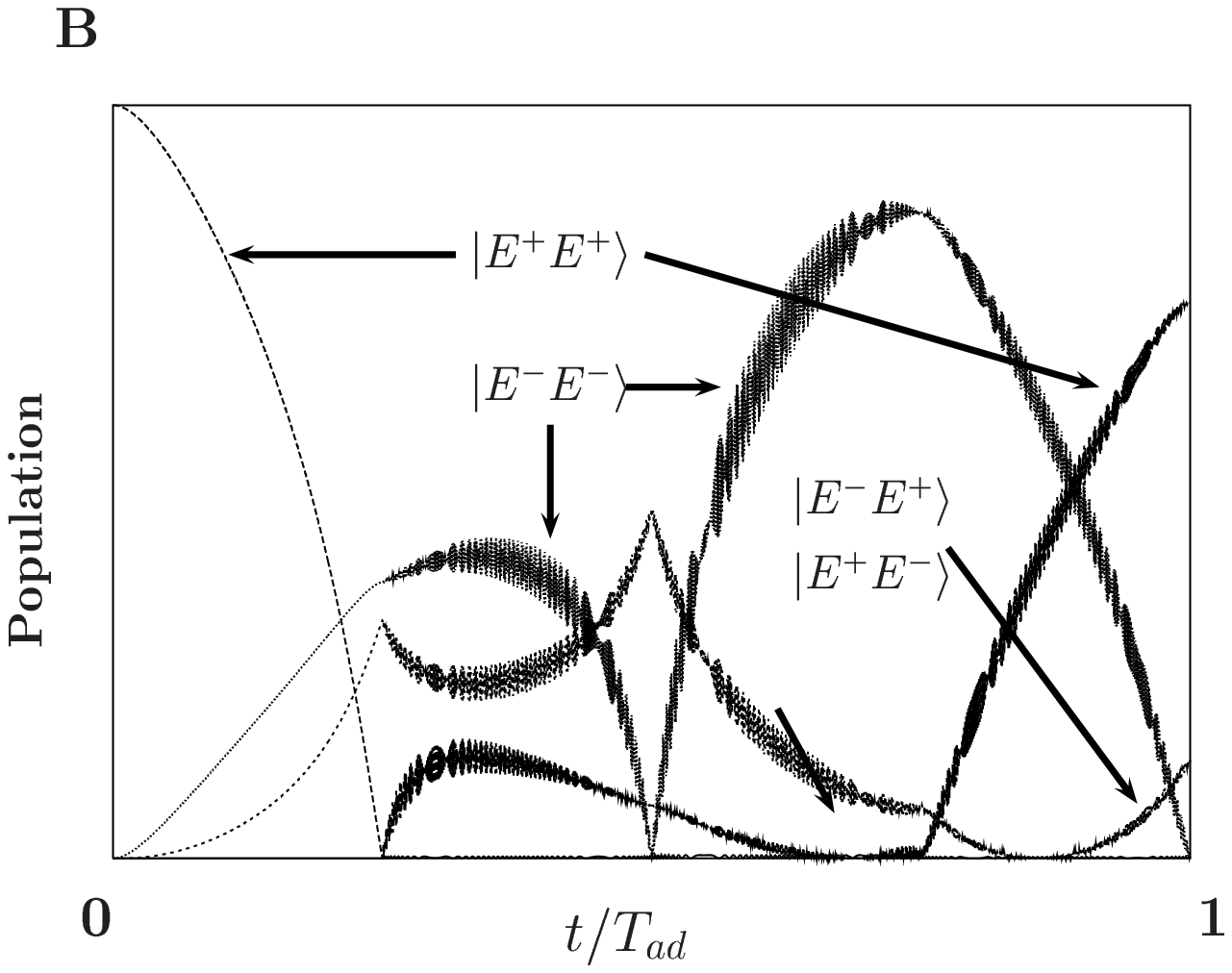}
  \caption{\label{fig:two_qubits} (A) Production of a bi-exciton state and 
isolation of the  $|E^i\rangle^{\otimes 2}-|G\rangle^{\otimes 2} $ space 
with the detuned lasers. (B) Simulated population evolution for the two 
qubit phase gate.}
\end{figure}

To show how the two-photon process happens we solved numerically
the Schroedinger equation for a four-level system
($|E E \rangle $, $|E G \rangle $, $|G E \rangle $,
$|G G \rangle $).
In Fig. \ref{fig:two_qubits} (A) we show the population evolution of the
states; the Rabi oscillation between $|E E \rangle $ and $|GG \rangle $
are evident and the states $|E G \rangle $, $|G E \rangle $ are
never populated. In order to fulfill the perturbation condition we choose 
$\delta/ \Omega = 25$.  

We have another degree of freedom in our system: the polarization.
Shining the dot with circular or linear polarization and we will obtain
$|G G \rangle  \rightarrow |E^i E^j\rangle $ ($i,j=+,-,0$) 
and can reproduce the scheme with polarized excitons.
The general Hamiltonian for the two-photon process is 
(in interaction representation)

\begin{equation}
   H_{int}=-\frac{2 \hbar^2}{\delta}
   \sum_{i,j=+,-,0} (\tilde{\Omega}_{i}\tilde{\Omega}_{j} 
   e^{i (\phi_i + \phi_j)}
   |E^{i} E^{j}\rangle \langle GG| + h.c.)
   \label{eq:two_ph}
\end{equation}

The total two-exciton space has dimension nine but we can restrict to 
four dimension space turning off two laser with the same polarization 
(i.e. $-$ or $0$) and because of this situation 
the scheme is slightly different for the one proposed in the other papers .
In ref. \cite{paper1} we show how to construct a phase gate; 
turning on the $+$ and $0$ lasers and modulating them to simulate the 
evolution in gate $1$ we were able to obtain  the geometrical operator
$U_3 =  exp(i \phi |E^+\rangle \langle E^+| ^{\otimes 2})$.
We can decouple the logical states with negative energy 
but we still need four laser (two with $+$ polarization and two with 
$0$ polarization) to produce a loop in the 
$|E^+\rangle ^{\otimes 2}- |E^0\rangle ^{\otimes 2} $ space.
The $+$ and $0$ lasers must be resonant with the two-exciton transition,
but in this scheme we also produce not-logical states 
$|E^+E^0\rangle$ and  $|E^0E^+\rangle$ since have the same energy 
($\omega_i^1 +\omega_j^1 =2E+\delta$).
Then we have a bigger dark space with dimension three
and the scheme is not directly repeated.
A detailed calculation of the 
dark states is given in appendix \ref{app:dark_states}.

Now we show how to construct another two-qubit geometrical operator 
with the same scheme.
Since in general an adiabatic loop will produce a superposition
of {\it all} the dark states, we change laser 
polarization ($0 \rightarrow -$) in order that the system can evolves in 
the logical space.
We note that the space is big enough to produce non-trivial transformation 
even without the {\it ancilla} qubits.

We choose the single laser Rabi frequencies in order to have 
$\Omega^{++} = \Omega \seno$,   
$\Omega^{--} = \Omega \coseno$, 
$\Omega^{+-} = \Omega \sqrt{|\seno~\coseno}|$ 
and $0 \leq \theta \leq 4 \pi$.
The dark state are 

\begin{eqnarray}
 |D_1\rangle&=& \coseno|++\rangle - \seno |--\rangle  \nonumber \\ 
 |D_2\rangle&=&\frac{1}{\sqrt{2}} (|+-\rangle - |-+\rangle) \nonumber\\
 |D_3\rangle&=&
 \sqrt{\frac{ |\sin \theta|}{1+ |\sin \theta|}} 
 (\seno |++\rangle + \coseno |--\rangle) \nonumber \\
 &-&\frac{1}{\sqrt{2(1+ |\sin\theta|)}} (|+-\rangle + |-+\rangle)
\end{eqnarray}

The associated connection is

\begin{equation}
A_\theta = 
\left ( \begin{array}{ccc}
             0       & 1/2 \sqrt{\frac{|\sin \theta|}{1+|\sin \theta|}} \\  
       -1/2 \sqrt{\frac{|\sin \theta|}{1+|\sin \theta|}}
     &  0 
             \end{array} \right )       
\label{eq:connection} 
\end{equation}  

Of course, for different values of $\theta~$, $[A_\theta, A_{\theta^\prime}]=0$
and we can calculate the loop integral and then the holonomy.
From numerical calculation we have 
$\alpha = \oint 1/2 \sqrt{\frac{|\sin \theta|}{1+|\sin \theta|}}d 
\theta = 
\int_0^{4 \pi} 1/2 \sqrt{\frac{|\sin \theta|}{1+|\sin \theta|}} d \theta
= 3.6806$ we have for the holonomic operator 

\begin{equation}
U = e^{-\alpha \sigma_x} = 
\left ( \begin{array}{ccc}
             \cos \alpha       & -\sin \alpha \\ 
             \sin \alpha      &  \cos \alpha 
             \end{array} \right )        
\end{equation}    

We write explicitly the final state using 
$|D_1 (4\pi)\rangle =|E^+\rangle ^{\otimes 2}$ and 
$|D_2 (4\pi)\rangle =1/\sqrt{2}(|E^+E^-\rangle + |E^-E^+\rangle) $

\begin{eqnarray}
  U |E^+\rangle^{\otimes 2}& = & \cos \alpha |E^+\rangle^{\otimes 2} \nonumber \\
  & -& \frac{\sin \alpha}{\sqrt{2}} (|E^+E^-\rangle + |E^-E^+\rangle)
\end{eqnarray}

This is an entangling gate, and then we have another non-trivial 
gate. In Fig. \ref{fig:two_qubits} we show the numerical simulation 
obtained solving the Schroedinger equation. It is difficult to follow the 
evolution of the states because of the number of the states populated 
during the evolution and because of the mixing of them.
Moreover it can be noted 
that the $|GG\rangle$ state never appears in the evolution, the 
$|E^-\rangle^{\otimes 2}$ state is not present at the end of the evolution 
and the final state is a superposition of $|E^+\rangle^{\otimes 2}$ and 
(symmetrically) $|E^+E^-\rangle - |E^-E^+\rangle$.

\section{conclusions }

In summary, we have shown that geometrical gates can be implemented in
quantum dots with optical control. 
We use  polarized excitons to encode logical information
 and we have been  able to construct a universal set of geometrical quantum gates.
The  biexcitonic shift due to exciton-exciton dipole coupling 
is exploited to implement two qubit gates.
Numerical simulations clearly suggest that one should  able 
 to apply several holonomic gates within the decoherence time.

Even though the fault-tolerance features of this geometrical approach
has not been completely  clarified so far (see e.g., \cite{critic}),
HQC surely provides, on the one hand,  a sort of an intermediate step towards topological quantum computing
and on the other hand, it is a natural arena in which explore  fascinating  quantum phenomena.

Finally we hope that the theoretical 
investigations here present will be effective in stimulating 
novel experimental activity in the field of coherent phenomena in semiconductor nanostructures. 
\acknowledgements{Funding by European Union project  TOPQIP Project,
(Contract IST-2001-39215) is gratefully aknowledged}

\appendix
\section{two-photon process}
\label{app:two_ph}

Here we show how two-photon process may occur in our system.
Let's consider two coupled quantum dots. The energy level spacing in this case
is different, in fact the presence  of an exciton in
one of them (e.g. in dot b) produces a shift in the energy level of the other
(e.g. dot a) from $E$ to $E + \delta/2$.
We have the following Hamiltonian

\begin{eqnarray}
  H_0 = (2E+\delta) (|E\rangle \langle E|)^{\otimes 2} + \nonumber \\
  E (|EG\rangle \langle EG|+|GE\rangle \langle GE|)
\end{eqnarray}

Using two lasers with frequencies $\omega= (E +\delta/2)$ ($\hbar=1$) 
the interaction
Hamiltonian is (we explicitly take into account the time dependence
$\Omega_i = \tilde{\Omega}_i e^{-i \omega t}$ from (\ref{eq:omega}))

\begin{equation}
  H_{int} = -\hbar \sum_{i=1,2} (\tilde{\Omega}_i e^{-i \omega t} |E_i\rangle \langle G_i|
  + \tilde{\Omega}_{i}^{*} e^{i \omega t} |G_i\rangle \langle E_i|)
  \label{eq:real_h}
\end{equation}

The effective Hamiltonian for the process is
(the apex $2$ indicate that is a second order process)

\begin{equation}
        H_{int}^{(2)}= - \hbar \tilde{\Omega} e^{-i \tilde{\omega} t} |E\rangle \langle G|^{\otimes 2}
        + h.c.
        \label{eq:fenom}
\end{equation}

where $\tilde{\omega}=2 \omega$ is the frequency that produces the transition
between $|GG\rangle$ and $|EE\rangle$.
There are four possible states ($|GG\rangle$ ,$|EG\rangle $ ,$|GE\rangle $ ,
$|EE\rangle$); let the
initial state be $|GG\rangle $ and we want to know the amplitude coefficient
for the $|GG\rangle \rightarrow |EE\rangle $ (Fig. \ref{fig:two_ph}).
To do this we use the interaction picture

\begin{eqnarray}
  \langle i| e^{i H_0 t/\hbar} H_{int} e^{-i H_0 t/\hbar} |j\rangle  = \nonumber \\
  e^{i (\omega_{i}-\omega_{j}) t}  e^{\pm i \omega t} \langle i|\tilde{H}_{int}|j\rangle
\end{eqnarray}

(the matrix element $\langle i|\tilde{H}_{int}|j\rangle $ is time independent)
with the initial conditions
$|\psi(0)\rangle  = |GG\rangle $ ($|\psi(t)\rangle  = \sum c_{ij}(t) |i j\rangle $
, $i,j=E,G$),  $\omega^{\prime} = \omega_{EE} - \omega_{m}$ and
$\omega^{\prime \prime} = \omega_{m} $, with perturbation theory to the
second order we get ($|m\rangle $'s are the intermediate states $|EG\rangle $ and $|GE\rangle $
with energy $E=\hbar \omega_{m}$)

\begin{eqnarray}
  c_{EE}^{(2)}(t) & = & (-\frac{i}{\hbar})^2 \sum_{m} \int_{0}^{t} d \tau_1
  \langle EE|\tilde{H}_{int}|m\rangle
  e^{i(\omega^{\prime} - \omega) \tau_1} \nonumber \\
  & & \int_{0}^{\tau_1} d \tau_2 \langle m|\tilde{H}_{int}|GG\rangle
  e^{i(\omega^{\prime \prime} - \omega) \tau_2}
\end{eqnarray}

Using $\omega^{\prime \prime}+\omega^{\prime}-2 \omega = 0$
, performing the double integration we get

\begin{eqnarray}
  c_{EE}^{(2)}(t) = & (-\frac{i}{\hbar})^2 \sum_{m}\langle EE|\tilde{H}_{int}|m\rangle
  \langle m|\tilde{H}_{int}|GG\rangle  \nonumber \\
  &  \frac{1}{i(\omega^{\prime \prime} - \omega)}
  (t-\frac{e^{i(\omega^{\prime} - \omega)t}-1}{i(\omega^{\prime} -
  \omega)})
  \label{eq:single_ph}
\end{eqnarray}

The term ${1 - e^{i(\omega^{\prime} - \omega)t}}{i(\omega^{\prime} -\omega)}$
oscillates, so the leading term is proportional to $t$

\begin{equation}
  c_{EE}^{(2)}(t) \approx  \frac{i}{\hbar^2}
  \sum_{m}\frac{\langle EE|\tilde{H}_{int}|m\rangle
  \langle m|\tilde{H}_{int}|GG\rangle }{\omega^{\prime \prime} -  \omega} t
\end{equation}

Now we go back to the second order Hamiltonian (\ref{eq:fenom})
(two-photon process) and calculate the evolution
($\Delta \omega = \omega_{EE} = 2 \omega$ and $\Delta \omega - \tilde{\omega}
=0$ )

\begin{eqnarray}
  c_{EE}^{(2)}
   & =  & - \frac{i}{\hbar} \int_{0}^{t} d t_1
   \langle EE|\tilde{H}_{int}^{(2)}|GG\rangle  e^{i(\Delta\omega - \tilde{\omega}) t_1} =
   \nonumber\\
   & & = - \frac{i}{\hbar}
   \langle EE|\tilde{H}_{int}^{(2)}|GG\rangle  \int_{0}^{t} d t_1 = \nonumber \\
   & & = - \frac{i}{\hbar} (-\hbar \tilde{\Omega})  t
   =  i \tilde{\Omega} t
   \label{eq:double_ph}
\end{eqnarray}

The two $c_{EE}^{(2)}$'s are calculated to the same order, so
using (\ref{eq:single_ph}) and (\ref{eq:double_ph})

\begin{equation}
  \tilde{\Omega}=\frac{1}{\hbar^2} \sum_{m} \frac{\langle EE|\tilde{H}_{int}|m\rangle
  \langle m|\tilde{H}_{int}|GG\rangle }{\omega^{\prime \prime} -
    \omega}
\end{equation}

In our system
\begin{eqnarray}
  \langle EE|\tilde{H}_{int}|EG\rangle \langle EG|\tilde{H}_{int}|GG\rangle = \nonumber \\
  = \langle EE|\tilde{H}_{int}|GE\rangle \langle GE|\tilde{H}_{int}|GG\rangle = \nonumber \\
  = \hbar^2 \tilde{\Omega}_1 \tilde{\Omega}_2
\end{eqnarray}

and we have the Rabi frequency for the two-photon process as
function of the single photon process ($\omega^{\prime \prime}_{m} -\omega =
\delta / \hbar$).
\begin{equation}
  \tilde{\Omega}= \frac{2 \hbar \tilde{\Omega}_1\tilde{ \Omega}_2}{\delta}
\end{equation}

We take into account the two exciton production for $E^{+}$ and $E^{0}$,
and choose : $\tilde{\Omega}_{1i}=\tilde{\Omega}_{i}$,
 $\tilde{\Omega}_{2i}=\tilde{\Omega}_{i} e^{i \varphi_{i}}$ with $i=+,0$.
The phenomenological Hamiltonian (\ref{eq:fenom})     became

\begin{eqnarray}
   H_{int} = - \frac{2 \hbar^2}{\delta}
   \tilde{\Omega}^2 e^{i \varphi} |E\rangle
   \langle G|^{\otimes 2}
   + h.c.
\end{eqnarray}

\section{Holonomic structure  of the  two-photon process}
\label{app:dark_states}

To explicitly calculate the dark state of Hamiltonian (\ref{eq:two_ph})
we change notation and include the phase in the definition of Rabi 
frequencies 
$\Omega^{ij} = \tilde{\Omega}_{i}\tilde{\Omega}_{j} e^{i (\phi_i + \phi_j)}$,
rewrite the Hamiltonian taking account of production of the same spin 
excitons ($i=j$) and choose the loop in order to have 
symmetric Rabi frequencies $\Omega^{ij}= \Omega^{ji}$, we obtain 
(with $|E^i\rangle = |i\rangle$) :

\begin{eqnarray}
   H_{int} &=& - \frac{2 \hbar^2}{\delta}
   ((\Omega^{++})^* |++\rangle + (\Omega^{jj})^* |jj\rangle \nonumber \\
   &+& (\Omega^{+j})^* (|+j\rangle + |j+\rangle) 
   \langle GG| 
   + h.c.
   \label{eq:ham_app}
\end{eqnarray}

where we can take $j=0,-$ to implement to different gates since we reduce 
the dark space and work with just two polarized excitons.

In addition to the decoupled states which do not appear in \ref{eq:ham_app},

we have three dark states ($\Omega^2 = |\Omega^{++}|^2 + |\Omega^{jj}|^2$) 

\begin{eqnarray}
 |D_1\rangle&=&\frac{(\Omega^{jj})^*|++\rangle - (\Omega^{++})^* |jj\rangle }{\Omega} \nonumber \\ 
 |D_2\rangle&=&\frac{1}{\sqrt{2}} (|+j\rangle - |j+\rangle) \nonumber\\
 |D_3\rangle&=&\frac{1}{\Omega \sqrt{|\Omega^{ij}|^2+ \Omega^2/2}} 
 [(\Omega^{ij})^*(\Omega^{++}|++\rangle + \Omega^{jj}|jj\rangle \nonumber)\\
 &-& \frac{\Omega^2}{2} (|+j\rangle + |j+\rangle)]
\label{eq:generic_dark}
\end{eqnarray}

Now we can explicitly calculate some connection for particular loops.
we choose $j=0$ and for the laser Rabi frequencies
$\Omega^+_i = \sqrt{\Omega \sin(\theta/2)}~exp(i\varphi/2)$, 
$\Omega^0_i = \sqrt{\Omega \cos(\theta/2)}$ ($i=1,2$ is the dot index)
and we use a loop in the $\theta$ and $\phi$ plane similar to the one 
in figure \ref{fig:theta-phi} ($0 \leq \theta \leq \pi$ and  
$0 \leq \phi \leq \pi/2$); then we have for the effective Rabi frequencies 

\begin{eqnarray}
  \Omega^{++} &=& \Omega \seno e^{i\varphi} \nonumber \\ 
  \Omega^{00} &=& \Omega \coseno \nonumber \\
  \Omega^{+0} &=& \Omega \sqrt{\seno~\coseno}~
    exp(i\varphi/2) 
\end{eqnarray}

The dark states in \ref{eq:generic_dark} explicitly take the form 

\begin{eqnarray}
 |D_1\rangle&=& \coseno|++\rangle - \seno e^{-i\varphi}  ||00\rangle  \nonumber \\ 
 |D_2\rangle&=&\frac{1}{\sqrt{2}} (|+0\rangle - |0+\rangle) \nonumber\\
 |D_3\rangle&=&
 \sqrt{\frac{ \sin \theta}{1+ \sin \theta}} 
 (\seno e^{i\varphi/2}|++\rangle + \coseno e^{-i\varphi/2} |00\rangle) \nonumber \\
 &-&\frac{1}{\sqrt{2(1+ \sin \theta)}} (|+0\rangle + |0+\rangle)
\end{eqnarray}

The connection associated is 

\begin{equation}
A_\theta = 
\left ( \begin{array}{ccc}
             0       & 1/2 \sqrt{\frac{\sin \theta}{2+\sin \theta}}~e^{i \varphi /2}  \\  
       -1/2 \sqrt{\frac{\sin \theta}{1+\sin \theta}}~e^{-i \varphi/2}
     &  0 
             \end{array} \right )        
\end{equation}  

\begin{equation}
A_\varphi = 
\left ( \begin{array}{cc}
             -i \sin^2 \frac{\theta}{2} & i/2 \sqrt{\frac{\sin \theta}{2+\sin \theta}} \sin \theta ~e^{i \varphi /2}   \\  
             i/2 \sqrt{\frac{\sin \theta}{1+\sin \theta}}\sin \theta ~e^{-i \varphi /2}    &  i/2 \frac{\sin \theta}{ 1+ \sin \theta } 
             \end{array} \right )        
\end{equation}  

The holonomic operator cannot be analytically calculated
because the connections do not commute. 
Then we calculated it with computer simulations by discretization of the 
loop in the parameter space.

\end{document}